\documentclass[twocolumn,pra,aps,showpacs,superscriptaddress]{revtex4}
\usepackage{xspace,amsmath,amsfonts,amsthm,amssymb,amsbsy,graphicx,color}
\begin{document}

\newtheorem{theo}{Theorem}
\newtheorem{lemma}{Lemma} 
%@twocolumnfalse\endcsname

\title{Quantum communication via a continuously monitored dual spin chain }

\author{Kosuke Shizume}

\affiliation{Institute of Library and Information Science, University of Tsukuba, 1-2 Kasuga, Tsukuba, Ibaraki 305-8550, Japan}

\author{Kurt Jacobs}

\affiliation{Department of Physics, University of Massachusetts at Boston, 100 Morrissey Blvd, Boston, MA 02125, USA}

\affiliation{Hearne Institute for Theoretical Physics, Department of Physics and Astronomy, Louisiana State University, Baton Rouge, LA 70803, USA} 

\author{Daniel Burgarth}

\affiliation{Department of Physics \& Astronomy, University College London, Gower St., London WC1E 6BT, UK}

\author{Sougato Bose}

\affiliation{Department of Physics \& Astronomy, University College London, Gower St., London WC1E 6BT, UK}

\begin{abstract}
We analyze a recent protocol for the transmission of quantum states via a dual spin chain [Burgarth and Bose, Phys. Rev. A {\bf 71}, 052315 (2005)] under the constraint that the receiver's measurement strength is finite. That is, we consider the channel where the ideal,  instantaneous and complete von Neumann measurements are replaced with a more realistic continuous measurement. We show that for optimal performance the measurement strength must be ``tuned'' to the channel spin-spin coupling, and once this is done, one is able to achieve a similar transmission rate to that obtained with ideal  measurements. The spin chain protocol thus remains effective under measurement constraints.   
\end{abstract}

\maketitle

\section{Introduction}
The task of reliably transferring quantum states between spatially separated systems is an important challenge in quantum information processing, and it has been shown that such transfers can be achieved using quantum spin chains~\cite{Bose03}. Such quantum channels avoid the need for interfacing solid-state qubits with their mobile counterparts, and may therefore prove important, especially for relatively short-distance communication. A number of schemes have been proposed that provide for high-fidelity transfer~\cite{Subrahmanyam04,Christandl04,Albanese04,Osborne04,Verstraete04,Verstraete04b,Jin04,Yung04,Amico04,Plenio05,Giovannetti05,Burgarth05,Haselgrove05,Wojcik05,Burgarth05b,Burgarth05c,Giovannetti06,Burgarth06,Lyakhov06}, and one of the most practical is the use of a dual spin chain~\cite{Burgarth05}. This provides heralded success, requires only a simple encoding, and maintains the advantage that the spin chain only need be manipulated at its end points. 

We will refer to each spin in the two chains as a {\em node} of the chain. 
To implement the communication protocol, the sender encodes the state across the two end-nodes of the dual chain at his end, and after a specified time has elapsed the receiver makes a projection measurement on the two end-nodes at his end. On obtaining the desired measurement outcome the state is successfully transferred to the receiver's end nodes and can be copied to a single qubit in the receiver's possession with local operations. The protocol as it stands thus requires the use of instantaneous projection measurements.  Such measurements are usually assumed in constructions of quantum communication protocols for simplicity. In this work we move beyond the simple assumption of instant projection measurements, and consider the implementation of the dual-spin-chain channel when instead the receiver continually monitors her end-nodes.  We do this for two reasons. The first is so that we can compare the performance of the channel with two very different kinds of measurements, and in particular to investigate whether continuous measurement might provide a more efficient means of implementing the channel. The second reason is that it is likely that in practical implementations of the transfer scheme, the time-scale of measurement will be similar to that of the spin-spin interactions that mediate the state-transfer along the chain. In this case an instantaneous projection measurement no longer serves as a good approximation; one must take into account the dynamics of the measurement process, and this necessitates a treatment using the machinery of continuous measurement~\cite{Brun02,JacobsSteck06}. 

Before we consider including continuous measurement in the implementation of the spin-chain channel, we now describe the operation of this channel. The channel consists of two parallel spin chains, each of which is an isotropic Heisenberg chain with Hamiltonian 
\begin{equation}
   H_i = B \sum_{n=1}^{N-1} \sigma_z^{(i,n)} + 
           J \sum_{n=1}^{N-1} \vec{\sigma}^{(i,n)} \cdot \vec{\sigma}^{(i,n+1)} ,
          \label{eq::H}
\end{equation}   
where the indices $i=1,2$ and $n=1,\ldots,N$ label, respectively, the chain and the position of the spin in the chain, with $n=1$ being the spins at the sender and $n=N$ being the spins at the receiver. The vector operator $\vec{\sigma} = (\sigma_x,\sigma_y,\sigma_z)$ is the vector of Pauli spin operators. A magnetic field is applied in the z-direction, the strength of which is given by $B$, and $J$ is the strength of the coupling between the spins. The chain is taken to be ferro-magnetic so that $J<0$. 

We will denote the spin states of the spins in each chain by $|0\rangle^{(i)}_n$ and $|1\rangle^{(i)}_n$, where $\sigma_z |0\rangle^{(i)}_n = - |0\rangle^{(i)}_n$ so that the ground state of each chain is 
\begin{equation}
  |{\mathbf 0}\rangle^{(i)} = |0\rangle^{(i)}_1|0\rangle^{(i)}_2 \cdots |0\rangle^{(i)}_N. 
\end{equation} 
To implement the transmission protocol the sender encodes a qubit state $|\psi\rangle = \alpha |0\rangle + \beta |1\rangle$ in the two end spins by placing them in the joint state   
\begin{equation}
   |\psi_0\rangle = \alpha |0\rangle^{(1)}_1|1\rangle^{(2)}_1 + \beta |1\rangle^{(1)}_1
                                       |0\rangle^{(2)}_1 .
   \label{eq::code}
\end{equation}   
The rest of the spins in each chain remain in their ground state. Denoting the state $|1\rangle$ for each of the spins as an ``excitation'' (a natural terminology), there is now exactly one excitation in the two chains. The two spin chains are then allowed to evolve under their Hamiltonians. Note that these Hamiltonians preserve the number of excitations in each chain, and we can therefore think of the excitation in both chains as moving along them (and spreading out across them) . If the coded qubit was at the $n^{\mbox{\scriptsize\em th}}$ node of the dual chain, then the state would be
\begin{equation}
   |\psi_n\rangle = \alpha |0\rangle^{(1)}_n|1\rangle^{(2)}_n + \beta |1\rangle^{(1)}_n
                                        |0\rangle^{(2)}_n .
\end{equation}   
Because the Hamiltonians of the two chains are the same, we can in fact also think of the coding state as moving along the dual chain. That is, we can write the state of the dual chain at time $t$ as  
\begin{equation}
  |\Psi(t)\rangle =  \sum_{n=1}^N c_n(t) |\psi_n\rangle 
\end{equation}   
for some complex coefficients $c_n(t)$. 
 
After waiting a time $\tau$ the receiver makes a parity measurement on the two spins that for the node at her end of the chain. This measurement projects the two spins into the subspace spanned by the ``even'' states $\{|0\rangle |0\rangle,|1\rangle|1\rangle\}$ or the odd states $\{|0\rangle|1\rangle,|1\rangle|0\rangle\}$. Because the coded qubit lies entirely within the odd state space, if the receiver obtains the result ``odd'', then she knows the state now lies at the end node, and has therefore been successfully transferred. If she obtains the result "even", then she has projected the end node onto the state $|0\rangle |0\rangle$, and knows that the qubit state remains in the rest of the dual chain. She can then repeat the parity measurement as many times as she wants, at regular or irregular time intervals, until she obtains the result ``odd'' and achieves a successful transfer. 

In the next section we show how the treatment of the transfer protocol is modified when the receiver makes a continuous measurement on the end-node. In Section~\ref{Results} we present our results, and in Section~\ref{Conclusion} conclude with a discussion and directions for future work. 

\section{The dual-chain protocol with continuous measurement}

When an observer continuously extracts information from a quantum system regarding an observable $X$, the dynamics induced in the system is given by the {\em stochastic master equation} (SME) 
\begin{equation}
  d\rho = -k[X,[X,\rho]] dt + \sqrt{2\eta k}(X\rho + \rho X - 2\langle X \rangle \rho ) dW 
  \label{eq::sme}
\end{equation}
Here $\rho$ is the density matrix of the system, $k$ is a constant, often called the {\em measurement strength}~\cite{DJJ}, that determines the rate at which information is extracted about $X$, and $\eta$ is the measurement efficiency.  The equation is {\em stochastic} because it is driven by the random Gaussian increment of Wiener noise, $dW$, which satisfies the Ito calculus relation $dW^2 = dt$~\cite{WienerIntroPaper}. The stochastic evolution is a direct result of the continuous random stream of measurement results obtained by the observer, usually referred to as the {\em measurement record}, $r(t)$, and which is given by 
\begin{equation}
   dr = \langle X \rangle dt + \frac{dW}{\sqrt{8\eta k}}
\end{equation} 
Readily accessible introductions to continuous quantum measurement, in which this equation is derived, are given in References~\cite{Brun02,JacobsSteck06}. 

If the system is subject to no other dynamics, or equally well posses a Hamiltonian that commutes with $X$ (so that $X$ is a QND observable~\cite{JLB07}), then the continuous measurement will project the system onto one of the eigenstates of $X$ as $t\rightarrow\infty$. The rate at which this occurs is proportional to $k (\Delta \lambda)^2$, where $\Delta \lambda$ is the average difference between adjacent eigenstates of $X$; this projection is essentially complete when $t\gg 1/[k(\Delta \lambda)^2]$~\cite{JacobsSteck06,JK}. Thus if $k$ is much larger than any other dynamical timescale in the system, then the continuous measurement described by Eq.(\ref{eq::sme}) is the usual instantaneous von Neumann measurement projecting the system onto an eigenstate of $X$. 

Returning to the dual-chain communication protocol described in the introduction, we see that if the receiver continually monitors the parity of the her end-nodes, then the dynamics of the channel is augmented by Eq.(\ref{eq::sme}). In this case $X$ is the operator 
\begin{equation}
   X = {\cal P}_{\mbox{\scriptsize even}} - {\cal P}_{\mbox{\scriptsize odd}} ,
\end{equation}
where ${\cal P}_{\mbox{\scriptsize even}}$ is the projector onto the even subspace of the  receivers end-nodes, and ${\cal P}_{\mbox{\scriptsize odd}}$ onto the odd subspace.

%To implement the protocol, the sender must prepare his end-nodes in the qubit code state Eq.(\ref{eq::code}). To ensure the receivers measurement does not interfere with the receivers preparation, we can do one of two things. During the preparation either the receiver can switch off her measurement, or the sender can decouple his end-nodes from the chain, whichever is easier. Once the sender has coded the state, the dual-chain evolves under the Hamiltonians of the two chains (Eq.(\ref{eq::H})) and the SME Eq.(\ref{eq::sme}). 

Recall that when the receiver uses an instantaneous measurement, it is the measurement result that tells her whether the coded state has arrived at her end-nodes. This is also true now. In the present case the receiver uses the measurement record to obtain the Wiener increment $dW$ and tracks the evolution of the dual chain by integrating the SME. This allows the receiver to determine the state of her end-nodes as the chain evolves. The receiver can thus calculate the overlap of the state of her end-nodes with the odd subspace. This overlap is a lower bound on the fidelity of the state present at the end nodes with the state being sent. The receiver waits until the overlap reaches a predetermined threshold, and then transfers the received state from the end-nodes to her local registers. 

The time that the receiver will have to wait for the overlap to reach the threshold will vary each time the protocol is implemented. This is because it depends on the random stream of the receiver's measurement results. In order to evaluate the performance of the transmission protocol, we must therefore simulate the relevant stochastic master equation for many different realizations of the measurement Wiener noise. The evolution of the system for each realization is referred to as a {\em trajectory}. Fortunately, it is not necessary to simulate the evolution in the full space of the spin chains, because the problem can be reduced to the sector in which there is only one excitation in the chains. This means that the effective size of the Hilbert space is $N$, where $N$ is the length of each chain. Even so, the simulations are sufficiently numerically intensive that the use of parallel supercomputers is invaluable. 

%%%%%%%%%%%%%%%%%%%%%%%%%%%%%%%%%%%%%%%%%%%%%%%%
\begin{figure}[t]
  \begin{center}
     \includegraphics[width=1.0\hsize]{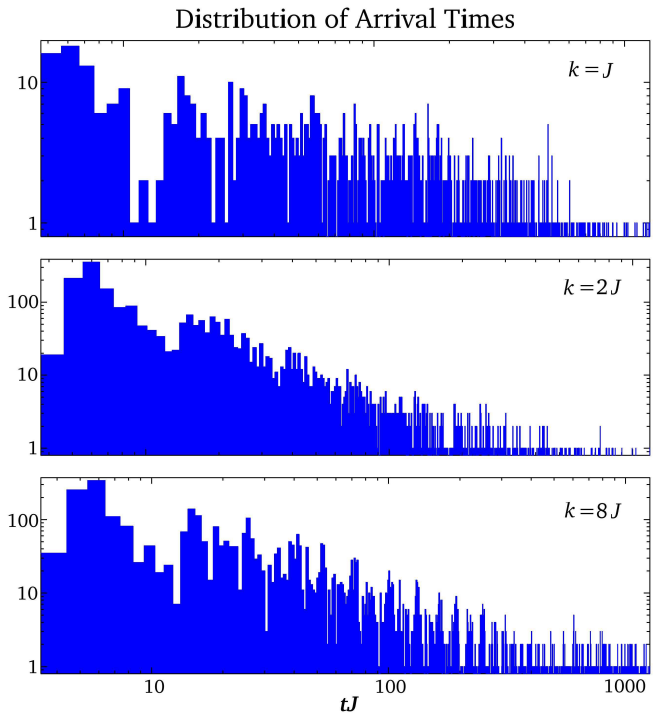}  % for arXiv submission
  \end{center}
  \vspace{-5mm}
  \caption
         { Histograms on a log-log scale of the ``arrival times'' of the quantum 
           state, for a fidelity threshold of $0.99$, and three values of the measurement 
           strength $k$. 
	\label{fig:dist1}}
\end{figure}
%%%%%%%%%%%%%%%%%%%%%%%%%%%%%%%%%%%%%%%%%%%%%%%%

%%%%%%%%%%%%%%%%%%%%%%%%%%%%%%%%%%%%%%%%%%%%%%%%
\begin{figure}[t]
  \begin{center}
     \includegraphics[width=1.0\hsize]{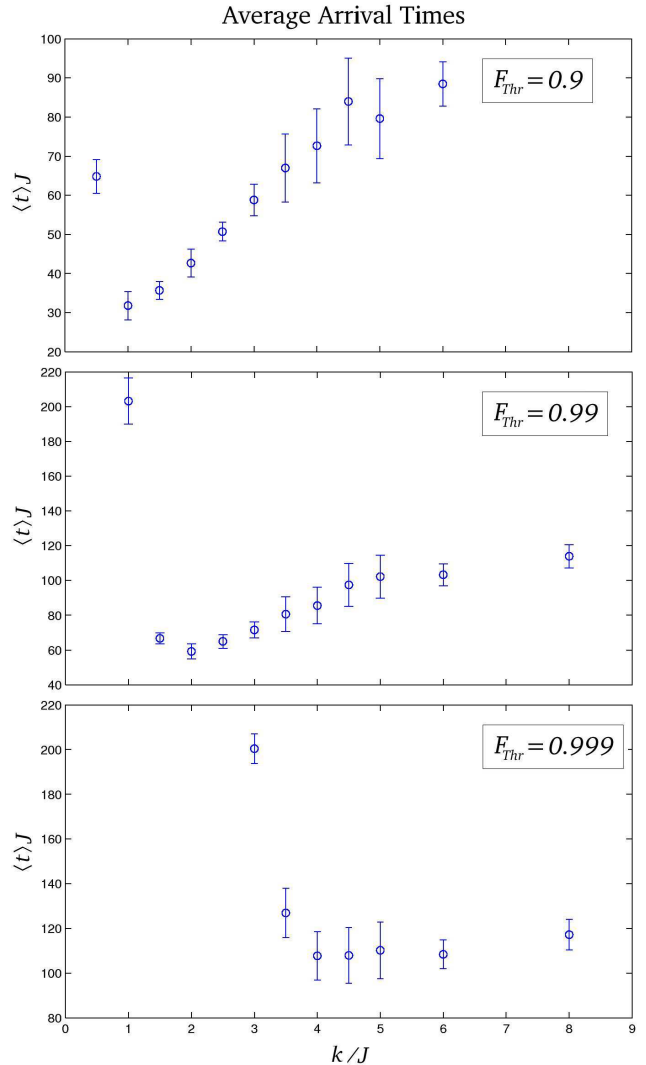}   % for arXiv submission
  \end{center}
  \vspace{-5mm}
  \caption
         {Here we plot the average time taken to transfer the state from the sender to the receiver as a function of the measurement strength $k$, and for three values of the fidelity, $F_{\mbox{\scriptsize\it Thr}}$, that is required for the transmitted state. 
	\label{fig:avt}}
\end{figure}
%%%%%%%%%%%%%%%%%%%%%%%%%%%%%%%%%%%%%%%%%%%%%%%%

The one free parameter in the protocol we have described is the strength of the measurement (in relation to the interaction strength $J$), and we expect the choice of this strength to have a significant effect upon the performance of the channel. In fact, one expects there to be an optimal measurement strength due to the quantum Zeno effect~\cite{Misra77,Home97}. That is, if the receiver measures her end-nodes too strongly, they will tend to remain in the original state. On the other hand, if she measures them to weakly, then she will be less likely to project the end-nodes into a state well within the odd subspace, and therefore have to wait longer for an effective transfer. One of the goals of our simulations is therefore be to investigate the dependence of the average transfer time on the measurement strength.  

\section{Results}
\label{Results} 

To calculate the performance of the dual-chain channel under a continuous measurement by the receiver, we must simulate the stochastic master equation given by Eq.(\ref{eq::sme}), with the inclusion of the Hamiltonian for each chain (Eq.(\ref{eq::H})). That is, 
\begin{eqnarray}
  d\rho & = & -i[(H_1 + H_2),\rho] dt  -k[X,[X,\rho]] dt  \nonumber \\ 
           &    & + \sqrt{2\eta k}(X\rho + \rho X - 2\langle X \rangle \rho ) dW 
  \label{eq::sme2}
\end{eqnarray}
Each realization of the noise represents a possible evolution of the channel under the observers measurement, and the time that the receiver must wait to obtain the state will vary from realization to realization. To obtain information about the average behavior of the channel we must therefore simulate Eq.(\ref{eq::sme2}) for a large enough number of realizations of the measurement process to obtain reasonable statistics. We set $\hbar=1$ and choose the interaction strength of the chain to be $J=1$. We thus quote all rate constants in units of $J$, and measure time in units of $1/J$. We choose the two spin chains to have a length of $N=10$ spins, and assume that the observers measurement is efficient, so that $\eta = 1$. We perform the numerical simulations using a simple half-order Newton method, and find that a time-step of $10^{-4} J^{-1}$ is sufficient. We perform simulations for a range of values of $k$, and for between $1024$ and $4096$ trajectories for each value. Since both chains are identical, the excitations in each chain travel at the same rate, and this is independent of the state being sent. In our simulations we choose the state to be that with $\alpha=\beta = 1/\sqrt{2}$. In order to decide when the state has ``arrived'' at the receiver's end-nodes, we choose a threshold for the fidelity between the original coding state, and the state at the receiver's end-nodes. We will refer to the time at which this fidelity crosses the threshold as the ``arrival time''. The more stringent the fidelity threshold the longer the receiver will have to wait. In the following analysis we will use the threshold values $F_{\mbox{\scriptsize\it Thr}} = 0.9$, $0.99$ and $0.999$.  

To begin we calculate histograms of the arrival times for the different values of $k$. This shows us that the distribution of arrival times has {\em very} long tails. In Fig.~\ref{fig:dist1} we show these histograms for $k=J, 2J$ and $8J$ with a threshold value of $F_{\mbox{\scriptsize\it Thr}} = 0.99$. Because of the long tails we display the histograms on a log-log scale. As an example of the long tails, for $k=2J$ and $F_{\mbox{\scriptsize\it Thr}} = 0.99$, while the mode of the arrival time is approximately $6 J^{-1}$, the average arrival time is $\langle t \rangle = 59 \pm 4 J^{-1}$, and $2\%$ of the trajectories give an arrival time longer than $440 J^{-1}$ (the latter being over $70$ times the mode). Examining the histogram for $k=2J$ we see that the distribution of arrival times falls approximately as a power law between $t=20 J^{-1}$ and $100 J^{-1}$, but appears to be sub-power law for longer times.  

We next calculate the average time of arrival for the transferred state as a function of the measurement strength, and for three values of the fidelity threshold. These results are displayed in Fig.~\ref{fig:avt}. We see that for all values of the threshold there is an optimal value of the measurement strength for which the average time of arrival is minimal. The monotonic rise in the waiting time above this optimal value is the expected  manifestation of the quantum Zeno effect. As the threshold is increased, the optimal value of the measurement strength increases. For the thresholds $0.9$, $0.99$ and $0.999$ the optimal values of $k$ are approximately $J$, $2J$ and $4J$, respectively. This behavior is not unreasonable, since one requires a stronger measurement to more fully project the system into the desired subspace.  We also see that there is a sharp rise in the waiting time when the measurement is weak, and this becomes sharper as the threshold is increased.  The sharp rise appears to imply that there is a threshold-like behavior in the ability of the measurement to localize the system sufficiently to the desired subspace;  when $k$ is below some critical value, then it is unlikely to ever localize the system to the desired subspace. The reason for the increase in the sharpness of the rise with decreasing fidelity threshold is not so easy to postulate, but may simply be due to the fact that the optimal value of $k$ is closer to the $k$-threshold when the fidelity threshold is lower. The optimal average arrival time  necessarily increases with fidelity, and is approximately $30 J^{-1}$, $60 J^{-1}$ and $110 J^{-1}$, respectively,  for the three values of the fidelity.

%%%%%%%%%%%%%%%%%%%%%%%%%%%%%%%%%%%%%%%%%%%%%%%%
\begin{figure}[t]
  \begin{center}
     \includegraphics[width=1.0\hsize]{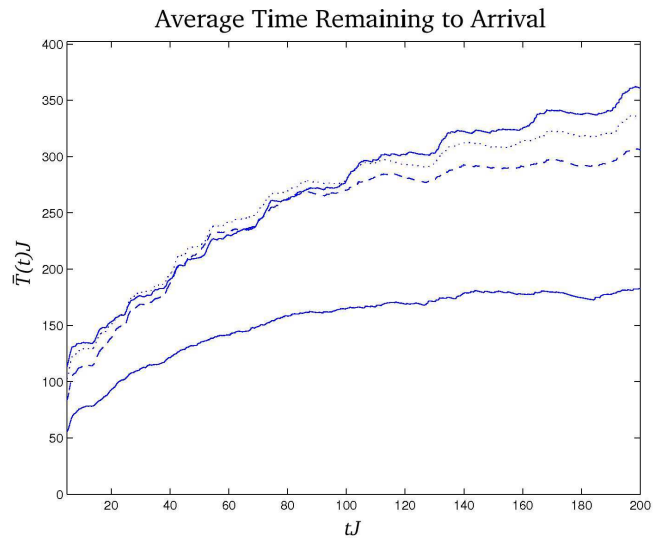}    % for arXiv submission 
  \end{center}
  \vspace{-5mm}
  \caption
         {Here we plot the average remaining time,  $\bar{T}$, that receiver expects 
         to have to wait for the state to arrive given that it has not yet arrived at time 
         $t$, as a function of $t$. We plot this is for a 
         threshold of $F_{\mbox{\scriptsize\it Thr}} = 0.99$ and four values of the 
         measurement strength $k$. Upper solid line: $k=8J$; dashed line: $k=6J$; 
         dotted line: $k=4J$; lower solid line: $k=2J$. 
	\label{fig:tleft}}
\end{figure}
%%%%%%%%%%%%%%%%%%%%%%%%%%%%%%%%%%%%%%%%%%%%%%%%

There is another quantity of potential importance, and that is the remaining time that the receiver expects to have to wait for the state to arrive, given that the state has not arrived at the current time $t$. This provides the receiver with information as to whether he should ``cut his losses'' and reset the channel so that the transmission can be repeated; if this average  time-remaining-to-arrival is increasing with time, and if it is longer than the sum of the time required to reset the channel and the average time-to-arrival at $t=0$, then the channel should be reset and the state re-sent. We denote the average time-remaining-to-arrival as $\bar{T}(t)$, and estimate this by averaging the arrival times in the interval $(t, \infty)$ obtained in our simulations, and subtracting $t$. We plot this for $F_{\mbox{\scriptsize\it Thr}} = 0.99$ and $k/J = 2,4,6$ and $8$ in Fig.~\ref{fig:tleft}. We see that the expected time remaining until arrival {\em increases} with time. That is, the longer the receiver waits for the state to arrive, the longer he expects to have to wait. In a given implementation it is likely that the time required to reset the channel will be longer than the average transmission time, since resetting will involve measurements by both sender and reciever. However, if the reset time were much smaller than the average  transmission time, then the receiver should only wait a short time before resetting the channel and requesting a resend, since the expected transmission time increases as the receiver waits. 

Finally, we compare the results for the protocol employing continuous measurement with that analyzed in~\cite{Burgarth05}, in which instantaneous von Neumann measurements are made by the receiver at judicious times. Specifically, in the dual-chain protocol introduced by two of us~\cite{Burgarth05}, the receiver waits until the probability that the end-node will be found in the coding state is high, and then makes a measurement to project the node into that state (that is, into the odd parity subspace). If the measurement fails, then the receiver repeats the process. The total failure probability drops exponentially as a function of the number of iterations.  For the case of a dual spin chain of length $N=10$, ones finds that the average time for a successful transmission is $ \langle t \rangle = 32.3 J^{-1}$. In this case the transmission is obtained with perfect fidelity, since the  von Neumann measurements are assumed to be perfect. We can compare this transmission time to those obtained with a continuous measurement for diffferent values of required fidelity. For a fidelity of $0.9$ we have $\langle t \rangle = (32 \pm 4) J^{-1}$, for $0.99$ we have $\langle t \rangle = (59 \pm 4) J^{-1}$, and $0.999$ gives  $\langle t \rangle = (108 \pm 11) J^{-1}$. These results are very encouraging; they show that the (more realistic) continuous measurement performs similarly to a sequence of von Neumann measurements. The only reduction in performance is an approximate factor of 2 increase in the transmission time for each factor of ten increase in the fidelity above a baseline of 0.9. 

\section{Conclusion}
\label{Conclusion}
We have analyzed the operation of a dual-spin chain quantum channel, in which the transfer of the quantum state is achieved when the receiver makes a measurement that projects his end nodes into a particular subspace. We have investigated the performance of the channel when the receiver replaces a sequence of ideal projection measurements by a continuous measurement. In this case the transmission rate is strongly dependent upon the strength of the measurement, with strong measurements inducing the quantum Zeno effect, and weak measurements reducing the likelihood of a definitive projection. Optimizing over the measurement strength, we find that the average rate of transmission is similar to that obtained with the original protocol. However, the transmission rate now depends upon the fidelity that is required of the transmitted state. For a fidelity of $0.9$ the rate of transmission is essentially the same as a channel with instant and perfect measurements, but increases as this fidelity is increased. This increase is not exorbitant however, as a fidelity of $0.999$ only decreases the average transmission rate by a factor of $4$.  

The above analysis raises a number of questions. The first is that, since the simulations are numerically intensive, we have only investigated the behavior for a chain of length $N=10$. Thus it remains an open question as to whether the behavior we have observed will be preserved in much longer chains. The second question has to do with the optimality of the protocol. For simplicity we assumed that the receiver kept his measurement on, and at the same strength, for the duration of the transmission, and optimized over this measurement strength.  This optimization therefore does not explore all the strategies that are available to the reciever. When the receiver makes instant projection measurements, he must wait between subsequent measurements in order to avoid the quantum Zeno effect which would block all transmission. Thus, it may well be optimal for an observer with a constrained measurement strength to similarly modulate the strength of his measurement.  Certainly we can see that if the receiver can make a stronger measurement than the optimal value found above, then he might well obtain a better result by using the maximal available strength and periodically turning the measurement off so as to beat the quantum Zeno effect. In addition, we note that in recent work Lyakhov and Bruder have shown that in the case of ideal measurements the transmission fidelity can be improved by modulating the interaction strength of the end nodes of a spin chain~\cite{Lyakhov06}, and this might also prove useful here. To take this idea to its limit, the globally optimal strategy is likely to involve a process of feedback control in which the receiver make his measurement strength, or the chain interaction strength, or both, dependent upon the measurement results at all earlier times~\cite{Jacobs06xb,Geremia04,Jacobs07}.  It remains an open question as to how much the receiver can improve upon the performance reported here by making the measurement strength time dependent, given a maximal value of this strength. The results we have presented indicate that this will also be a function of the desired fidelity. 

\section*{Acknowledgments} Part of the numerical results in this research were obtained using supercomputing resources at Information Synergy Center, Tohoku University. KJ acknowledges the support of The Hearne Institute for Theoretical Physics, The National Security Agency, The Army Research Office and The Disruptive Technologies Office.  DB acknowledges the support of the UK Engineering and Physical Sciences Research Council, Grant Nr. GR/S62796/01.

%\bibliographystyle{apsrev}
%\bibliography{report}

\end{document}